\documentclass[aps,prl,twocolumn,superscriptaddress,longbibliography]{revtex4-2}
\usepackage{graphicx}
\usepackage{amssymb,amsmath}
\usepackage{bm,bbold}
\usepackage{dcolumn}
\usepackage{float}
\usepackage[OT1]{fontenc} 
\usepackage{url}
\usepackage{mathrsfs}
\usepackage{slashed,comment}
\usepackage{color}
\usepackage{verbatim}
\usepackage{enumitem}
\usepackage{soul,physics}
\usepackage[driverfallback=dvipdfm]{hyperref}
\hypersetup{pdfpagemode=FullScreen,colorlinks=true,breaklinks,urlcolor=blue,linkcolor=blue,citecolor=blue}
\usepackage[english]{babel}
\usepackage{amsthm}
\usepackage{subfigure}

\newtheorem*{theorem}{Theorem}

\makeatletter
\renewcommand{\@fnsymbol}[1]{%
  \ensuremath{%
    \ifcase#1
    \or \dagger      
    \or \ddagger     
    \or \mathsection 
    \or \mathparagraph
    \else\@ctrerr
    \fi
  }%
}
\makeatother

\begin{document}
 
  \title{Exact Hilbert-space Ergodicity from Continuous Monitoring}

  \author{Yue Wu}
  \affiliation{Institute for Advanced Study, Tsinghua University, Beijing 100084, China}

  \author{Yuzhi Tong}
  \affiliation{Department of Physics \& State Key Laboratory of Surface Physics, Fudan University, Shanghai, 200438, China}
  \affiliation{Institute for Advanced Study, Tsinghua University, Beijing 100084, China}

  \author{Liang Mao}
  \thanks{Corresponding author: 
  \href{liangmao.physics@gmail.com}{{liangmao.physics@gmail.com}}}
  \affiliation{Institute for Advanced Study, Tsinghua University, Beijing 100084, China}

  \author{Pengfei Zhang}
  \thanks{Corresponding author: 
  \href{pengfeizhang.physics@gmail.com}{{pengfeizhang.physics@gmail.com}}}
  \affiliation{Department of Physics \& State Key Laboratory of Surface Physics, Fudan University, Shanghai, 200438, China}

  \date{\today}

  \begin{abstract}
  Quantum evolution is generally expected to drive a quantum many-body system toward equilibrium. This expectation is often justified by the Hilbert-space ergodicity of generic quantum dynamics, namely, the idea that pure-state evolution explores Hilbert space uniformly up to physical constraints. Such a statement can be made rigorous by requiring the associated state ensemble to form the Haar-random ensemble, or its more structured generalization, the Scrooge ensemble. In this Letter, we report the emergence of exact Hilbert-space ergodicity in a continuously monitored quantum many-body system. For any target density matrix $\sigma$, we construct a continuously monitored system for which we rigorously prove that the Scrooge ensemble of $\sigma$ is the unique late-time equilibrium distribution of quantum trajectories. Remarkably, this requires only that the jump operators in the monitoring form a deformed unitary 1-design, a seemingly much weaker condition than full ergodicity. We numerically demonstrate our predictions by simulating continuously monitored systems whose equilibrium states are thermal states. Our results establish a rigorous mechanism for the emergence of Hilbert-space ergodicity and provide a practical route for its investigation on quantum devices.

   \end{abstract}
  
  \maketitle

  \emph{ \color{blue}Introduction.--} 
  Late-time quantum evolution is generally expected to drive quantum many-body systems toward equilibrium, admitting universal descriptions such as thermal equilibrium~\cite{Borgonovi2016QuantumChaos,Gogolin2016Equilibration,Kaufman2016QuantumThermalization,Ueda2020QuantumEquilibration,Abanin2019Colloquium,Srednicki1994Chaos,Deutsch1991QuantumStatistical,Rigol2008Thermalization,Mori2018Thermalization,DAlessio2016QuantumChaos,Nandkishore2015ManyBodyLocalization}. This expectation is justified through the Hilbert-space ergodic property of quantum dynamics~\cite{Liu2026HierarchyHSE,PilatowskyCameo2025CriticallySlow,Shaw2025ExperimentalHSE,PilatowskyCameo2023CompleteHSE,PilatowskyCameo2024HSE,Mark2024MaximumEntropy}, which states that the quantum state is driven to explore the entire Hilbert space uniformly, up to physical constraints. This ergodic dynamics averages out fine-grained dynamical information, leaving only global information such as conserved charges or energy. When all structure is erased by the evolution, as in the case of fully chaotic quantum dynamics, Hilbert-space ergodicity can be defined rigorously by requiring the generated pure-state ensemble to form the Haar-random ensemble, the uniform ensemble on Hilbert space. The Haar-random ensemble has been used to describe generic chaotic quantum evolution~\cite{Dankert2009ExactApproximate2Designs,Brandao2016LocalRandomCircuits,Schuster2025ExtremelyLowDepth,Cui2025NearlyOptimalDepth,Cui2025RandomHamiltonianDynamics,Mao2025RandomUnitaries,Sun2026TwoChaoticHamiltonians,Dankert2005EfficientSimulation,Gross2007EvenlyDistributed,Ambainis2007QuantumTDesigns,Zhang2026MagicAugmentedClifford,Nakata2017EfficientPseudorandomness,Zhou2025SingleQuenchDesigns,Haah2024RandomPauliRotations,LaRacuente2026ShallowLowCommunication,page1993average,hayden2007black,cotler2017chaos,Roberts2017ChaosComplexityDesign,popescu2006entanglement,popescu2006entanglement,emerson2003pseudo,harrow2009random,nahum2017quantum,von2018operator}, and also has broad applications in quantum information~\cite{Dankert2009ExactApproximate2Designs,Huang2020ClassicalShadows,Arute2019QuantumSupremacy,Madsen2022ProgrammablePhotonicProcessor,Elben2023RandomizedMeasurementToolbox,Zhong2020QuantumAdvantagePhotons,Scott2008ProcessTomography2Designs,boixo2018characterizing,emerson2005scalable,aharonov2022quantum,huang2022quantum,chen2022exponential,ji2018pseudorandom,Schuster2025StrongRandomUnitaries,movassagh2023hardness,bouland2019complexity}.

  In practice, however, many systems cannot scramble toward Haar-random behavior due to physical constraints, yet nonetheless exhibit many hallmarks of scrambling and chaos. Based on earlier studies, recent progress suggests that the so-called Scrooge ensemble provides a universal description for these cases. The Scrooge ensemble is defined as a Haar-random ensemble deformed by a density matrix $\sigma$,
  \begin{equation}\label{eq:Scrooge_def}
  \mathcal{E}_{\mathrm{sc}}(\sigma)=\left\{D\langle\phi|\sigma |\phi\rangle d\phi,  \frac{\sqrt{\sigma}|\phi\rangle}{\sqrt{\langle\phi|\sigma|\phi\rangle}}\right\},
  \end{equation}
  where we denote a state ensemble by $\mathcal{E}=\{p_a,\ket{\psi_a}\}$, with $p_a$ being the probability (density) of $\ket{\psi_a}$. $d\phi$ is the Haar measure and $D$ is the Hilbert-space dimension. The Scrooge ensemble minimizes the accessible information among all ensembles with density matrix $\sigma$~\cite{Jozsa1994LowerBound}. Physically, it is the most random pure-state ensemble subject to the constraint that the average state equals $\sigma$. This suggests a natural definition of constrained Hilbert-space ergodicity: the ensemble generated by quantum dynamics approaches the Scrooge ensemble associated with $\sigma$. Indeed, recent progress has confirmed the ergodicity of time-independent Hamiltonian dynamics by identifying the generated temporal ensemble as a Scrooge ensemble~\cite{Mok2026NatureStingy,Linden2009QuantumMechanical,Mark2024MaximumEntropy}. The projective ensemble of small subsystems from generic quantum states has also been confirmed to form a Scrooge ensemble~\cite{Liu2026CoherenceInduced,Mok2026NatureStingy,Cotler2023EmergentDesigns,Ho2022ExactEmergent,Liu2024GaussianDeepThermalization,Mark2024MaximumEntropy,Ippoliti2023DynamicalPurification,Chang2025ChargeConserving,Claeys2022Biunitarity,Ippoliti2022SolvableDeepThermalization,Choi2023PreparingRandomStates,Feng2026ResourceLocalizability,mcginley2025scroogeensemblemanybodyquantum}.

  \begin{figure}[t]
    \centering
    \includegraphics[width=0.83\linewidth]{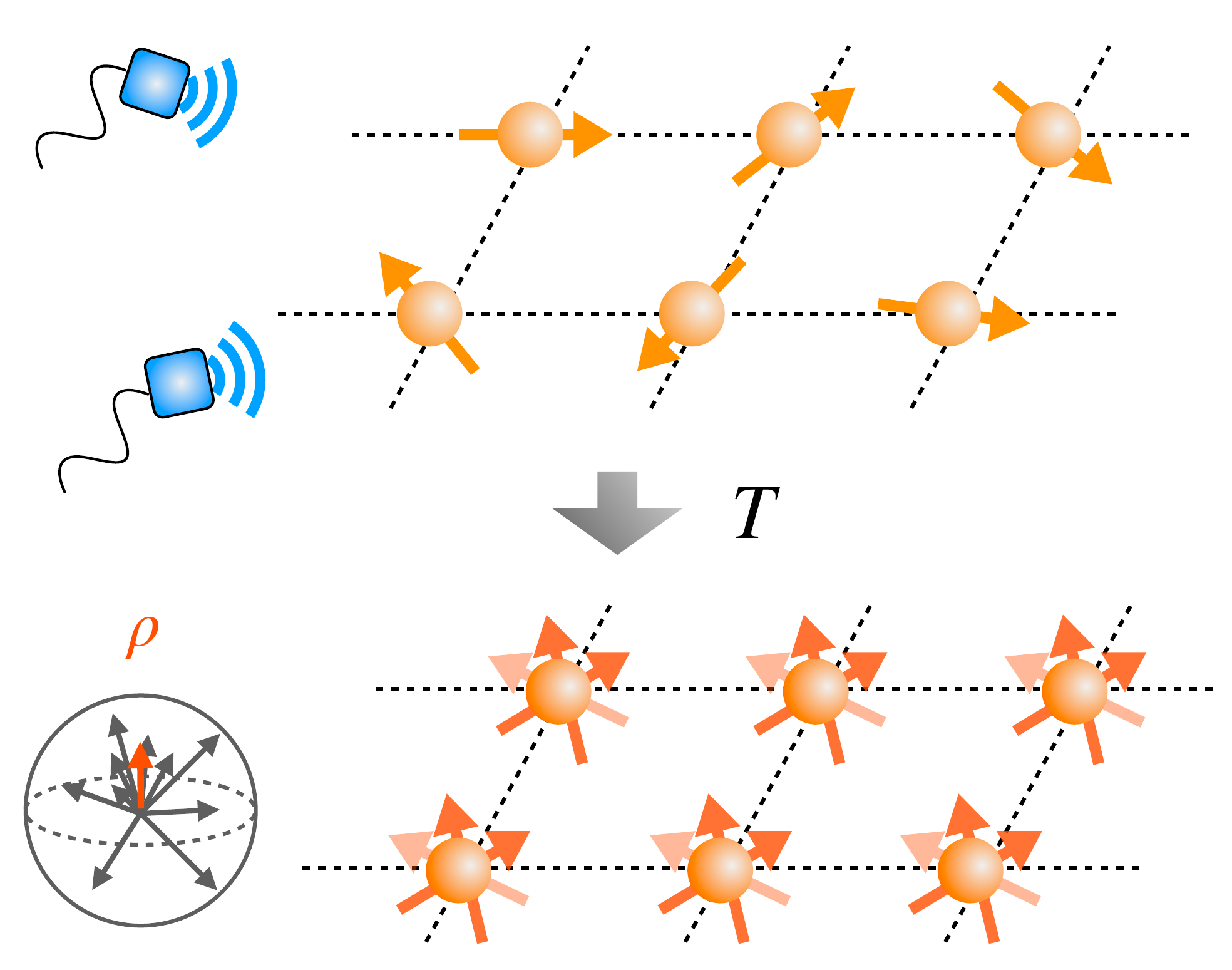}
    \caption{Schematic illustration of our main results. A quantum system is subjected to continuous measurements constructed from a prescribed density matrix. We prove that, after a sufficiently long time $T$, the ensemble of quantum trajectories converges to the exact Scrooge ensemble associated with the prescribed density matrix.  }
    \label{fig:schematics}
  \end{figure}

  In this Letter, we report a novel mechanism for realizing Hilbert-space ergodicity in the above sense. We show that, in quantum many-body systems subjected to continuous monitoring, Haar or Scrooge ensembles emerge at late times, as illustrated in Fig.~\ref{fig:schematics}. For any density matrix $\sigma$, we construct a corresponding continuous measurement system, for which we prove that $\mathcal{E}_{\mathrm{sc}}(\sigma)$ is \emph{exactly} the unique equilibrium distribution of quantum trajectories (where $\sigma=\mathbb{I}/D$ recovers the Haar case). Interestingly, this requires only that the jump operators in the monitoring form a deformed unitary 1-design, a much weaker condition than full ergodicity. Numerical simulations for different system sizes validate our theoretical predictions. Finally, we discuss experimental detection of the Scrooge ensemble in our systems.

  \emph{ \color{blue}Continuous monitoring.--} 
  Our scheme generates the Scrooge ensemble through continuous monitoring, implemented via successive weak measurements~\cite{Braginsky1992QuantumMeasurement,Carmichael1993OpenSystems,Barchielli2009QuantumTrajectories,Wiseman1996QuantumTrajectories,Daley2014QuantumTrajectories,Jacobs2006Straightforward,Habib2006EmergenceChaos,Bhattacharya2000ContinuousMeasurementChaos,Fuji2020MeasurementCriticality}. Such measurements extract an infinitesimal amount of information about the quantum state at each infinitesimal time step.
  At each step, we couple the system (in state $\ket{\psi}$) to an ancilla initialized in $\ket{0}$ through a unitary operator~\cite{Gross2018QubitModels}
  \begin{equation}
  V_\alpha(dt)=\exp\left[\sqrt{dt}\,\left(c_\alpha\otimes \ket{1}\bra{0}-c_\alpha^\dagger\otimes \ket{0}\bra{1}\right)\right].
  \end{equation}
  Here, $c_\alpha$ denotes a system operator whose explicit form will be specified later. After the coupling, the ancilla is measured in the $x$ basis, yielding a random outcome $r_\alpha=\pm1$. This measurement branches the input state into two post-selected quantum states, defining distinct quantum trajectories. For convenience, we introduce the random variable $dW_\alpha=r_\alpha\sqrt{dt}- O_\alpha(\psi) dt$ with $O_\alpha(\psi)= \langle \psi|c_\alpha+c_\alpha^\dagger|\psi\rangle$, which has zero mean and variance $dt$. The post-measurement quantum state is described by the stochastic Schrödinger equation
  \begin{equation}\label{eq:equation}
  d\ket{\psi}=\sum_\alpha A_\alpha(\psi)\ket{\psi}\,dt+\sum_\alpha B_\alpha(\psi)\ket{\psi}\,dW_\alpha,
  \end{equation}
  with $B_\alpha(\psi)=c_\alpha-\frac{1}{2}O_\alpha(\psi)$ and $A_\alpha(\psi)=-\frac{1}{2}c_\alpha^\dagger c_\alpha+\frac{1}{2}O_\alpha(\psi)c_\alpha-\frac{1}{8}O_\alpha(\psi)^2$. 
  Here, the sum over $\alpha$ accounts for a sequence of measurements characterized by different operators $c_\alpha$. In the limit $dt\rightarrow 0$, $dW_\alpha$ is the standard Wiener process, satisfying $dW_\alpha dW_\beta = \delta_{\alpha,\beta}dt$.
  By repeating the measurement process up to time $T$, we obtain a collection of quantum trajectories that define a random-state ensemble. 

  To facilitate ergodic dynamics with late-time density matrix $\sigma$, two requirements must be satisfied. First, the jump operators $c_\alpha$ must be sufficiently random to enable a complete exploration of Hilbert space. Second, the equilibrium density matrix should be $\sigma$. For the first requirement, one can take $c_\alpha\propto U_{\text{Haar}}$, where $U_{\text{Haar}}$ is a Haar-random unitary. For the second, it can be shown that the evolution of the density matrix $\rho(t)=\overline{|\psi(t)\rangle \langle \psi(t)|}$ is governed by the standard Lindbladian master equation
  \begin{align}
      \frac{d\rho}{dt}=\mathcal{L}[\rho]=\sum_\alpha \Big(
      c_\alpha \rho c^\dagger_\alpha-\frac{1}{2}\big\{c_\alpha^\dagger c_\alpha,\rho\big\}
      \Big).
  \end{align}
  It is therefore natural to choose 
  \begin{align}\label{eq:trial}
      c_\alpha\propto \sqrt{\sigma}U_{\text{Haar}},
  \end{align}
  so that the resulting master equation admits $\sigma$ as its stationary state and satisfies quantum detailed balance~\cite{alicki1976detailed,Agarwal1973Microreversibility}.

  Our finding is in fact stronger: instead of the Haar unitary in Eq.~\eqref{eq:trial}, a unitary 1-design suffices. 
  \begin{theorem}
    \label{thm:approx}
    In the long-time limit $T\to\infty$, the trajectories of the stochastic Schrödinger equation \eqref{eq:equation} exactly generate the Scrooge ensemble \eqref{eq:Scrooge_def} from an arbitrary initial pure-state ensemble, provided that the jump operators are chosen as $c_\alpha=\sqrt{\kappa D p_\alpha}\sqrt{\sigma}\,U_\alpha$, and the unitary ensemble $\mathcal{E}_U=\{p_\alpha,U_\alpha\}$ satisfies the conditions
    \begin{itemize}
    \item Unitary 1-design: $\sum_\alpha p_\alpha U_\alpha Q U_\alpha^\dagger=\frac{\text{Tr}[Q]}{D}\mathbb{I}$, for any operator $Q$ on the $D$-dimensional Hilbert space,
    \item Phase neutral: $\sum_\alpha p_\alpha U_\alpha\otimes U_\alpha=0$.
    \end{itemize}
  \end{theorem}

  We note that the phase-neutral condition can be naturally satisfied by assigning the same probability distribution to each unitary $U_\alpha$ and $i U_\alpha$. Consequently, the simplest choice satisfying both conditions in qubit systems is $\{P, iP\}$ with a uniform distribution, where $P$ runs over all Pauli strings~\cite{Dankert2005EfficientSimulation}. Moreover, it is straightforward to show that the master equation becomes
  \begin{equation}\label{eq:master}
  \frac{d\rho}{dt}=\mathcal{L}[\rho]=-\kappa(\rho-\sigma),
  \end{equation}
  which gives $\rho(t) = \sigma+(\rho-\sigma)e^{-\kappa t}$. Our construction, however, incorporates additional structure that ensures convergence of arbitrary moments to those of the Scrooge ensemble. It can therefore be regarded as a higher-order generalization of the quantum detailed balance condition.
  
  \emph{ \color{blue}Proof outline.--} We now sketch the proof, leaving details to the Supplemental Material. We begin by introducing a parameterization of the general wave function $\ket{\psi}$ and rewriting the Scrooge ensemble as an explicit probability distribution over the corresponding parameters. A convenient choice is the eigenbasis that diagonalizes the target density matrix $\sigma$, with eigenvalues $\{\lambda_i\}$ labeled by $i\in\{1,2,\cdots,D\}$. We first assume all eigenvalues are nonzero, corresponding to a full-rank density matrix. In this basis, we parametrize the wavefunction as $\psi_i=\sqrt{y_i}e^{i\theta_i}$. For the Scrooge ensemble, the distributions of $y_i$ and $\theta_i$ can be determined from those of the Haar ensemble. Expanding a Haar-random state $\ket{\phi}$ in the same basis as $\phi_i=\sqrt{x_i}e^{i\delta_i}$, the Haar ensemble corresponds to a uniform distribution of both $x_i$ and $\delta_i$ subject to the constraint $\sum_i x_i=1$. The definition \eqref{eq:Scrooge_def} then gives $y_i=\lambda_i x_i/s(\bm{x})$ and $\theta_i=\delta_i$, where $s(\bm{x})=\sum_{j=1}^{D}\lambda_j x_j$. It follows that the distribution of the phase variables $\theta_i$ is uniform, just as in the Haar ensemble. In contrast, the distribution of the amplitude variables $y_i$ is determined by the Jacobian of the change of variables together with the additional weighting factor $\langle\phi|\sigma|\phi\rangle$. This yields the classical probability distribution of the Scrooge ensemble:
  \begin{equation}
   p_{\text{sc}}(\bm{y},\bm{\varphi})= C\Bigg(\sum_{j=1}^D\frac{y_j}{\lambda_j}\Bigg)^{-(D+1)}.
  \end{equation}
  Here, $C$ is a normalization constant. We introduce the $2(D-1)$ independent variables $\bm{y}=(y_1,\ldots,y_{D-1})$ and $\bm{\varphi}=(\theta_1-\theta_D,\ldots,\theta_{D-1}-\theta_D)$, reflecting the constraint $\sum_i y_i=1$ and the redundancy of the overall phase.
  
  Next, we rewrite the stochastic Schrödinger equation \eqref{eq:equation} as an evolution equation for these independent classical variables. For the amplitude variables $y_i(t)$, the evolution takes the form 
  \begin{equation}\label{eq:diff}
  dy_i=\kappa(\lambda_i-y_i) dt+\sum_\alpha g_{i\alpha}(\psi)dW_\alpha.
  \end{equation}
  Here, we define $g_{i\alpha}(\psi)=\psi_i^*\eta_{\alpha,i}+\eta_{\alpha,i}^*\psi_i-O_{\alpha}y_i$ with $\ket{\eta_\alpha}=c_\alpha\ket{\psi}$. This equation describes a biased random walk in the parameter space of $y_i$. The first term represents a drift with velocity $v_i^{(a)}(\bm{y})=\kappa(\lambda_i-y_i)$ that drives $y_i$ toward $\lambda_i$, while the second term describes diffusion arising from stochastic measurement outcomes. In the theory of classical stochastic processes~\cite{Pavliotis2014StochasticProcesses}, the diffusion properties are encoded in the covariance matrix, defined by
  \begin{equation}
  \begin{aligned}
  D_{ij}^{(a)}(\bm{y})&=\overline{dy_i^{(d)}dy_j^{(d)}}/dt=\sum_\alpha g_{i\alpha}(\psi)g_{j\alpha}(\psi)\\&=2\kappa[\delta_{ij}\lambda_iy_i+(s(\bm{y})-\lambda_i-\lambda_j)y_iy_j].
  \end{aligned}
  \end{equation}
  Here, $dy_i^{(d)}$ denotes the diffusion term in Eq.~\eqref{eq:diff}, and both conditions on the jump operators have been used. 

  Similarly, the phase variables $\varphi_i$ exhibit zero drift and a finite diffusion matrix:
  \begin{equation}
  \begin{aligned}
  D_{ij}^{(\varphi)}(\bm{y})&=\frac{\kappa}{2}\left(\delta_{ij}\frac{\lambda_i}{y_i}+\frac{\lambda_D}{y_D}\right).
  \end{aligned}
  \end{equation}
  The diffusion matrix is independent of $\varphi_i$. Both this property and the vanishing drift follow from the invariance of the trajectory ensemble under independent phase shifts $\theta_i\rightarrow\theta_i+a_i$, as a direct consequence of the conditions imposed on the jump operators. Furthermore, the symmetry $\theta_i\rightarrow-\theta_i$ forbids mixed diffusion terms between $y_i$ and $\varphi_j$.

  The stochastic dynamics of the classical variables induces a corresponding evolution of their probability distribution $p(\bm{y},\bm{\varphi},t)$, governed by the standard Fokker-Planck equation~\cite{Pavliotis2014StochasticProcesses,altland2010condensed}. In terms of the drift vector and diffusion matrix, it reads
  \begin{equation}\label{eq:Fokker-Planck}
  \partial_t p=-\sum_\mu\partial_\mu\big(v_\mu\,p\big)+\frac{1}{2}\sum_{\mu\nu}\partial_\mu\partial_\nu \big(D_{\mu\nu}\,p\big).
  \end{equation}
  For conciseness, we introduce the $2(D-1)$-dimensional vector $z_\mu=(\bm{y},\bm{\varphi})$. The corresponding drift vector and diffusion matrix are given by
  \begin{equation}
  v_\mu=({v}^{(a)}_i,0),\ \ \ \ \ \ D_{\mu\nu}=
  \begin{pmatrix}
  D^{(a)}_{ij}&  0\\
  0&D_{ij}^{(\varphi)}
  \end{pmatrix}.
  \end{equation}
  This Fokker-Planck equation exactly describes the evolution of the ensemble of quantum trajectories at time $t$, where each point $z_\mu$ on the underlying manifold corresponds to a pure quantum state. 

  Establishing convergence to the Scrooge ensemble from an arbitrary initial state ensemble involves two steps. First, we show that the Scrooge distribution is a stationary solution of the Fokker-Planck equation by directly verifying that the right-hand side of Eq.~\eqref{eq:Fokker-Planck} vanishes when $p(\bm{y},\bm{\varphi})=p_{\mathrm{sc}}(\bm{y},\bm{\varphi})$. The calculation is straightforward, though somewhat lengthy. Second, we show that this steady state is unique. Mathematically, the underlying space of pure states forms the compact connected manifold $\mathbb{CP}^{D-1}$. Doob's theorem then guarantees uniqueness of the steady state, provided that the diffusion matrix $D_{\mu\nu}$ is uniformly positive definite for all $z_\mu$~\cite{KulikScheutzow2015,DaPratoZabczyk1996,SaloffCoste1992}. This condition can be verified directly, as elaborated in the Supplemental Material. Physically, positive definiteness implies that, at every pure state $\ket{\psi}$, the diffusion process has nonzero fluctuations in all local directions on the manifold. Consequently, each state is connected to nearby states, allowing the diffusion process to explore the entire Hilbert space and providing the physical intuition behind convergence to a unique steady-state distribution.

  Finally, although the preceding discussion assumes that $\sigma$ is full rank, the evolution of the first moment \eqref{eq:master} implies that all quantum trajectories are confined to the support of $\sigma$ in the long-time limit. The above analysis can therefore be repeated within this support. The corresponding Scrooge ensemble remains the unique steady state, and every initial ensemble, including those with support outside $\sigma$, converges to it asymptotically. This completes the proof of the theorem.

  \begin{figure}[t]
    \centering
    \includegraphics[width=0.99\linewidth]{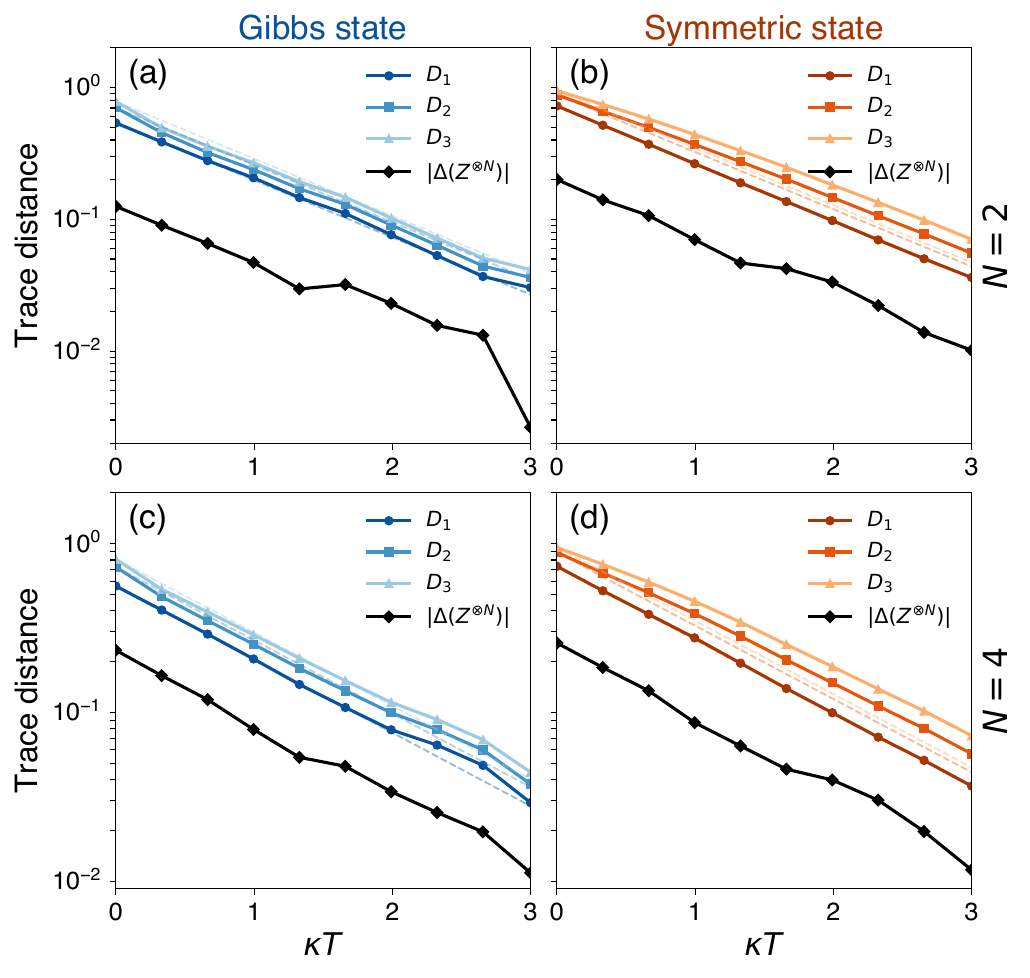}
    \caption{Numerical simulation of our protocol. (a)(c) The steady-state density matrix is the Gibbs state $\sigma_0$ of the quantum Ising model with $\beta J=2$ and $g/J=1/2$. (b)(d) The steady-state density matrix $\sigma_+$ is $\sigma_0$ projected onto the $+1$ eigenspace of the $\mathrm{Z}_2$ symmetry. In all four subfigures, we compute the distances $D_{1,2,3}$ from the Scrooge moments, as well as the difference $\Delta$ of the quantum-classical correlator $C^{QC}_2$ for $O_1=O_2=Z^{\otimes N}$, with its ideal value given by the expectation value of the Scrooge second moment. The dashed reference lines denote exponential decay with rate $\kappa$.
    In the simulation we fix $\kappa/J=1$, and compute the Scrooge moments via Monte Carlo sampling with 10000 samples. The empirical values of the moments of the trajectory ensemble and of the quantum-classical correlator are obtained by averaging over $N_a=5000$ trajectories.}
    \label{fig:numerics}
  \end{figure}

  \emph{ \color{blue}Numerics.--} We present numerical demonstrations of our protocol for qubit systems with different system sizes $N$. Specifically, we consider a target density matrix corresponding to the thermal state of the quantum Ising model, described by the Hamiltonian
  \begin{equation}
  H=-J\sum_{n=1}^{N-1}Z_nZ_{n+1}-g\sum_{n=1}^N X_n.
  \end{equation}
  The quantum Ising model possesses a global $Z_2$ symmetry generated by $R=\prod_n X_n$. This allows us to construct two distinct thermal density matrices, $\sigma_0=e^{-\beta H}/Z_0$ and $\sigma_+=\frac{(1+R)}{2}e^{-\beta H}/Z_+$, where $Z_0$ and $Z_+$ ensure the normalizations $\mathrm{tr}[\sigma_0]=\mathrm{tr}[\sigma_+]=1$. While $\sigma_0$ is supported on the entire Hilbert space and is therefore full rank, $\sigma_+$ is the thermal state projected onto the even-parity sector and thus has support only on a proper subspace. We consider both $\sigma_0$ and $\sigma_+$ as the prescribed density matrix for the target Scrooge ensemble. 

  To quantify the deviation between the numerically generated trajectory ensemble $\mathcal{E}_{\mathrm{qt}}=\{\ket{\psi_a}\}$ and the target Scrooge ensemble $\mathcal{E}_{\mathrm{sc}}$, we compare their $n$th moments through the distance
  \begin{equation}
  D_n\equiv \Bigg|\Bigg|
  \frac{1}{N_a}\sum_a
  \bigl(|\psi_a\rangle\langle\psi_a|\bigr)^{\otimes n}
  -\mathbb{E}_{\mathcal{E}_{\mathrm{sc}}}
  \bigl[
  (|\psi\rangle\langle\psi|)^{\otimes n}
  \bigr]
  \Bigg|\Bigg|_1,
  \end{equation}
  where $||\cdot||_1$ denotes the trace norm and $N_a$ is the number of sampled trajectories. The numerical results are presented in Fig.~\ref{fig:numerics}. We fix $\beta J=2$ and $g/J=1/2$, and consider systems with $N=2$ and $N=4$ qubits. The system is initialized in the fully polarized state $\ket{00\cdots0}$. The distance $D_n$ is plotted as a function of evolution time $\kappa T$ for $n\in\{1,2,3\}$. The results show that $D_n$ decays rapidly and approaches zero in the long-time limit, providing direct numerical evidence for convergence of the trajectory ensemble to the Scrooge ensemble.

  We further discuss relaxation toward the steady state. In general, relaxation to the stationary distribution is governed by the low-lying eigenmodes of the Fokker-Planck operator. Since different observables may couple to different eigenmodes, their apparent relaxation rates need not be identical. Remarkably, however, our numerical results indicate that $D_n$ for $n\in\{1,2,3\}$ all decay with the same relaxation rate $\kappa$, consistent with the prediction for the first moment from the Lindblad master equation~\eqref{eq:master}. Motivated by this observation, we conjecture that the relaxation of arbitrary moments is governed by the same leading eigenmode, exhibiting the same relaxation rate. If so, convergence to the Scrooge ensemble occurs on a timescale of order $\kappa^{-1}$, independent of both the moment order $n$ and the system size $N$.

  \emph{ \color{blue}Experimental considerations.--} Experimental verification of our scheme requires complete tomography of the quantum trajectories conditioned on the measurement record, which generally demands an exponential number of samples. Similar challenges arise in experimental studies of measurement-induced phase transitions~\cite{GoogleQuantumAI2023MeasurementInduced,Kamakari2025ScalableXEB,Li2023CrossEntropyMIPT} and strong-to-weak symmetry breaking~\cite{Wang2026ObservationSWSSB}. In those cases, the difficulty was circumvented by introducing quantum-classical cross correlators that can be measured efficiently. Motivated by these developments, we employ a similar approach to benchmark our protocol. 

  For simplicity, we focus on the second moment; the generalization to arbitrary moments is straightforward. On a quantum device, after obtaining a quantum trajectory $|\psi_a\rangle$, we perform a projective measurement of a Pauli operator $O_1$, yielding an outcome $o_1^Q(a)=\pm 1$ with probability $p_{\pm}=\langle\psi_a|\frac{1\pm O_1}{2}|\psi_a\rangle$. We then use knowledge of the trajectory $a$ to simulate the dynamics on a classical computer and compute the classical expectation $O_2^C(a)=\langle \psi_a |O_2|\psi_a\rangle$. The quantum-classical correlator is then
  \begin{equation}
  \begin{aligned}
  C^{QC}_2&\equiv\frac{1}{N_a}\sum_a o_1^Q(a) O_2^C(a)\\&\overset{N_a\rightarrow \infty}{=} \mathbb{E}_{\mathcal{E}_{\mathrm{qt}}}\,
  \text{tr} \left[O_1\otimes O_2
  (|\psi\rangle\langle\psi|)^{\otimes 2}\right].
  \end{aligned}
  \end{equation}
  By computing the difference $\Delta$ between $C^{QC}_2$ for sufficiently large $N_a$ and the Scrooge ensemble prediction for the observable $O_1\otimes O_2$ on the replicated Hilbert space, we obtain an experimentally accessible quantity that diagnoses the emergence of the Scrooge ensemble. We numerically simulate the decay of $\Delta$ for $O_1=O_2=Z^{\otimes N}$.
  The result is also shown in Fig.~\ref{fig:numerics}, demonstrating the experimental relevance of our scheme.

  \emph{ \color{blue}Discussions.--} In this Letter, we rigorously establish exact Hilbert-space ergodicity under an arbitrary prescribed density matrix via continuous measurements. Remarkably, to generate the maximally random ensemble of states at late times, it is not necessary for the jump operators themselves to be maximally random. Instead, it suffices that the unitaries entering the construction form a unitary 1-design and satisfy the phase-neutral condition. Since no special structure is required of the prescribed density matrix and the conditions on the jump operators do not depend on microscopic details, our results are broadly applicable across different quantum platforms. These findings show that higher-order statistical properties of quantum states can be systematically engineered via measurement design, and open a route toward realizing nontrivial random-state ensembles as steady states of quantum trajectories.

  Several interesting directions remain open. On the theoretical side, it would be valuable to establish analytically whether the observed relaxation rate $\kappa$ for higher moments reflects a deeper spectral property of the underlying Fokker-Planck operator. It is also natural to explore extensions of our approach to constrained random-state ensembles associated with additional conserved quantities or symmetries. On the practical side, the ability to prepare exact Scrooge ensembles may provide a useful resource for studying information hiding, state indistinguishability, and the universal statistical properties of complex quantum states on quantum devices.

  \textit{Acknowledgement.}
  We sincerely thank Hui Zhai for invaluable discussions. We thank ChatGPT 5.5 and Claude Opus 4.8w for supplying ideas for the proof, and helping write the code for Fig.~\ref{fig:numerics}. We have carefully written and verified the derivation, and are responsible for the correctness of this work.
  This project is supported by the Shanghai Rising-Star Program under grant number 24QA2700300, the NSFC under grant 12374477, 
  and the Xuemin Institute of Advanced Studies at Fudan University.

  \bibliography{ref.bib}

  \end{document}


\title{Supplementary materials: Exact Hilbert-space Ergodicity from Continuous Monitoring}

  \author{Yue Wu}
  \affiliation{Institute for Advanced Study, Tsinghua University, Beijing 100084, China}

  \author{Yuzhi Tong}
  \affiliation{Department of Physics \& State Key Laboratory of Surface Physics, Fudan University, Shanghai, 200438, China}
  \affiliation{Institute for Advanced Study, Tsinghua University, Beijing 100084, China}

  \author{Liang Mao}
  \affiliation{Institute for Advanced Study, Tsinghua University, Beijing 100084, China}

  \author{Pengfei Zhang}
  \affiliation{Department of Physics \& State Key Laboratory of Surface Physics, Fudan University, Shanghai, 200438, China}

  \date{\today}
  \maketitle

\section{Scrooge distribution}

In this section we derive the probability density of Scrooge ensemble. For any density matrix $\sigma$, we work in the eigenbasis of it. We can write $\sigma=\mathrm{diag}\{\lambda_1,\cdots\lambda_D\}$, with constraint $\sum_{i=1}^D\lambda_i=1$. We assume that $\sigma$ is full rank unless otherwise specified, so that
$\lambda_i>0$ for all $i=1,\ldots,D$. We expand a pure state \(\ket{\psi}\) on the eigenbasis of $\sigma$ and parametrize each component as $\psi_i=\sqrt{y_i}e^{i\theta_i}$. Then it can be described by $2D$ real numbers $\vec{y}=(y_1,\ldots,y_D)$ and $\vec{\theta}=(\theta_1,\ldots,\theta_D)$. Because of the constraint $\sum_iy_i=1$ and the redundancy of the overall phase, there are
$2(D-1)$ independent variables $\bm z=(\bm y,\bm\varphi)$, where $\bm y=(y_1,\ldots,y_{D-1})$ and  \(\bm \varphi=(\theta_1-\theta_D,\ldots,\theta_{D-1}-\theta
_D)\).

The Scrooge ensemble with average density matrix $\sigma$ can be constructed from Haar ensemble as:
\begin{equation}
    \mathcal{E}_{\mathrm{sc}}(\sigma)
  =
  \left\{
  D\bra{\phi}\sigma\ket{\phi}\,d\phi,\,
  \frac{\sqrt{\sigma}\ket{\phi}}
       {\sqrt{\bra{\phi}\sigma\ket{\phi}}}
  \right\}.
  \label{eq:scrooge-def-app}
\end{equation}
Let $\vec x=(x_1,\cdots ,x_D)$ and $\vec \delta=(\delta_1,\ldots,\delta_D)$ denote the probabilities and phases of a Haar random state $\ket{\phi}$, respectively, and $\bm x=(x_1,\cdots ,x_{D-1})$ denotes the independent variables. $\vec x$ obeys a uniform distribution on the probability simplex. $\vec \delta$ obeys a uniform distribution on $U(1)^{D}$. We can use this as a seed to generate a Scrooge state.
From the definition of Scrooge ensemble,
\begin{equation}
  y_i=\frac{\lambda_i x_i}{s(\bm x)},\qquad
  s(\bm x)=\sum_{k=1}^{D}\lambda_k x_k,\qquad \theta_i=\delta_i.
  \label{eq:y-from-x}
\end{equation}
Define
\begin{equation}
  H(\bm y)=\sum_{k=1}^{D}\frac{y_k}{\lambda_k},
\end{equation}
solving \eqref{eq:y-from-x} for \(x\) yields
\begin{equation}
  x_i=\frac{y_i/\lambda_i}{H(\bm y)},\qquad
  s(\bm x)=\frac{1}{H(\bm y)}.
  \label{eq:x-from-y}
\end{equation}
Since \(y_D\) is not an independent variable, it is determined by the normalization constraint as
\begin{equation}
y_D=1-\sum_{k=1}^{D-1}y_k.
\end{equation}
Therefore, \(H(y)\) can be expressed in terms of the independent variables as
\begin{align}
H(y)
&=\sum_{k=1}^{D}\frac{y_k}{\lambda_k} \nonumber\\
&=\sum_{k=1}^{D-1}\frac{y_k}{\lambda_k}
+\frac{1-\sum_{k=1}^{D-1}y_k}{\lambda_D}.
\end{align}
It follows that, for \(j=1,\ldots,D-1\),
\begin{equation}
\partial_j H
\equiv \frac{\partial H}{\partial y_j}
=\frac{1}{\lambda_j}-\frac{1}{\lambda_D}.
\end{equation}
Hence we obtain, for \(i,j=1,\ldots,D-1\),
\begin{align}
\frac{\partial x_i}{\partial y_j}
&=\frac{\delta_{ij}}{\lambda_i H(y)}
-\frac{y_i}{\lambda_i H(y)^2}\partial_j H \nonumber\\
&=\frac{1}{\lambda_i H(y)}
\left[
\delta_{ij}
-\frac{y_i}{H(y)}
\left(
\frac{1}{\lambda_j}-\frac{1}{\lambda_D}
\right)
\right].
\end{align}
Therefore the Jacobian matrix is
\begin{equation}
J
=
\operatorname{diag}\left(
\frac{1}{\lambda_1 H},
\dots,
\frac{1}{\lambda_{D-1} H}
\right)\cdot
\left[
I-u v^T
\right],
\end{equation}
where
\begin{equation}
u_i=\frac{y_i}{H},
\qquad
v_j=\frac{1}{\lambda_j}-\frac{1}{\lambda_D}.
\end{equation}

Using the matrix determinant rule,
\begin{equation}
\det(I-u v^T)=1-v^T u.
\end{equation}
Compute
\begin{equation}
v^T u
=
\frac{1}{H}
\sum_{j=1}^{D-1}
\left(
\frac{1}{\lambda_j}
-
\frac{1}{\lambda_D}
\right)y_j=\frac{H-\frac{1}{\lambda_D}}{H}.
\end{equation}
Hence
\begin{equation}
\det(I-u v^T)
=
1-\frac{H-\frac{1}{\lambda_D}}{H}
=
\frac{1}{\lambda_D H}.
\end{equation}
Thus
\begin{equation}
\det J
=
\left(
\prod_{i=1}^{D-1}
\frac{1}{\lambda_i H}
\right)
\frac{1}{\lambda_D H}
=
\frac{1}{\prod_{i=1}^D \lambda_i}
\frac{1}{H^D}.
\end{equation}

Including the Scrooge reweighting factor
$s(\bm x)=H(\bm y)^{-1}$, the transformed density is
\begin{equation}
p_{\mathrm{sc}}(\bm y,\bm\varphi)
\propto
s(\bm x)
\left|
\det\frac{\partial\bm x}{\partial\bm y}
\right|
\propto
H(\bm y)^{-(D+1)}.
\end{equation}
Hence
\begin{equation}
p_{\mathrm{sc}}(\bm y,\bm\varphi)
=
C H(\bm y)^{-(D+1)},
\end{equation}
where \(C\) is the normalization constant. Since the phases distribution is not altered, it keeps to be a uniform distribution.

\section{Fokker--Planck Equation Associated with the Stochastic Schr\"odinger Equation}

In this section, we derive the Fokker-Planck equations for parametriezed pure state from the Stochastic Schr\"odinger equation.
Throughout this section, we take the jump operators to be
\begin{equation}
c_\alpha
=
\sqrt{\kappa Dp_\alpha}\,
\sqrt{\sigma}\,U_\alpha,
\end{equation}
where the unitary ensemble
$\mathcal{E}_U=\{p_\alpha,U_\alpha\}$ satisfies
\begin{align}
\sum_\alpha p_\alpha
U_\alpha XU_\alpha^\dagger
&=
\frac{\operatorname{Tr}X}{D}\mathbb{I},\\
\sum_\alpha p_\alpha
U_\alpha\otimes U_\alpha
&=
0
\end{align}
for an arbitrary operator $X$. The first relation is the unitary
$1$-design condition, while the second is the phase-neutral condition.

Our goal is to derive the stochastic dynamics of the independent coordinates
\begin{equation}
\bm z=(\bm y,\bm \varphi),
\end{equation}
which parametrize the pure-state manifold, where
\begin{equation}
\bm y=(y_1,\ldots,y_{D-1}),
\qquad
\bm \varphi=(\theta_1-\theta_D,\ldots,\theta_{D-1}-\theta_D).
\end{equation}
We first consider the amplitude variables. Since we work in the eigenbasis
$\{|i\rangle\}$ of $\sigma$, the probabilities
\begin{equation}
y_i=|\psi_i|^2
\end{equation}
are precisely the diagonal matrix elements of the rank-one projector
\begin{equation}
P=|\psi\rangle\langle\psi|,
\qquad
y_i=\langle i|P|i\rangle.
\end{equation}
It is therefore convenient to first derive the stochastic equation for $P$.
The equations for the independent amplitude variables
$y_1,\ldots,y_{D-1}$ can then be obtained by taking the corresponding
diagonal matrix elements of $dP$.

Using the It\^o product rule, we have
\begin{align}
dP
&=
d|\psi\rangle\langle\psi|
+
|\psi\rangle d\langle\psi|
+
d|\psi\rangle d\langle\psi|.
\end{align}
Substituting
\begin{equation}
d|\psi\rangle
=
\sum_\alpha A_\alpha|\psi\rangle\,dt
+
\sum_\alpha B_\alpha|\psi\rangle\,dW_\alpha
\end{equation}
and using
\begin{equation}
dW_\alpha dW_\beta
=
\delta_{\alpha\beta}\,dt,
\end{equation}
we obtain
\begin{align}
dP
&=
\sum_\alpha
\left(
A_\alpha P
+
P A_\alpha^\dagger
+
B_\alpha P B_\alpha^\dagger
\right)dt
\nonumber\\
&\quad+
\sum_\alpha
\left(
B_\alpha P
+
P B_\alpha^\dagger
\right)dW_\alpha.
\end{align}
Substituting the explicit expressions
\begin{align}
B_\alpha
&=
c_\alpha-\frac{1}{2}O_\alpha,\\
A_\alpha
&=
-\frac{1}{2}c_\alpha^\dagger c_\alpha
+\frac{1}{2}O_\alpha c_\alpha
-\frac{1}{8}O_\alpha^2,
\end{align}
where
\begin{equation}
O_\alpha
=
\operatorname{Tr}
\left[
\left(c_\alpha+c_\alpha^\dagger\right)P
\right],
\end{equation}
gives
\begin{align}
dP
&=
\sum_\alpha
\left[
c_\alpha P c_\alpha^\dagger
-\frac{1}{2}
\left\{
c_\alpha^\dagger c_\alpha,P
\right\}
\right]dt
\nonumber\\
&\quad+
\sum_\alpha
\left[
c_\alpha P
+
P c_\alpha^\dagger
-
O_\alpha P
\right]dW_\alpha.
\end{align}

For later convenience, we separate the deterministic and stochastic
contributions by defining
\begin{align}
\mathcal{L}_{\alpha}(P)
&=
c_{\alpha}Pc_{\alpha}^{\dagger}
-\frac{1}{2}
\left\{
c_{\alpha}^{\dagger}c_{\alpha},P
\right\}, \\
M_{\alpha}(P)
&=
c_{\alpha}P
+
Pc_{\alpha}^{\dagger}
-
O_{\alpha}(P)P.
\end{align}
We also define the total Lindblad generator as
\begin{equation}
\mathcal{L}(P)
=
\sum_{\alpha}\mathcal{L}_{\alpha}(P).
\end{equation}
The stochastic equation for the projector can then be written compactly as
\begin{equation}
dP
=
\mathcal{L}(P)\,dt
+
\sum_{\alpha}M_{\alpha}(P)\,dW_{\alpha}.
\end{equation}
Here, $\mathcal{L}(P)$ determines the deterministic drift of the state,
whereas $M_{\alpha}(P)$ describes the stochastic backaction associated with
measurement channel $\alpha$.

Taking the
$i$th diagonal matrix element of the stochastic equation for $P$ gives
\begin{align}
dy_i
&=
\langle i|\mathcal{L}(P)|i\rangle\,dt
+
\sum_{\alpha}
\langle i|M_{\alpha}(P)|i\rangle\,dW_{\alpha}.
\end{align}
We evaluate the deterministic and stochastic contributions separately.

For the deterministic contribution, 
the first term in the Lindblad generator is
\begin{align}
\sum_{\alpha}
\langle i|c_{\alpha}Pc_{\alpha}^{\dagger}|i\rangle
&=
\kappa D
\sum_{\alpha}p_{\alpha}
\langle i|
\sqrt{\sigma}\,
U_{\alpha}PU_{\alpha}^{\dagger}
\sqrt{\sigma}
|i\rangle
\nonumber\\
&=
\kappa D\lambda_i
\sum_{\alpha}p_{\alpha}
\langle i|U_{\alpha}PU_{\alpha}^{\dagger}|i\rangle.
\end{align}

Using the unitary $1$-design condition,
\begin{equation}
\sum_{\alpha}p_{\alpha}
U_{\alpha}PU_{\alpha}^{\dagger}
=
\frac{\operatorname{Tr}P}{D}\mathbb{I}
=
\frac{1}{D}\mathbb{I},
\end{equation}
we obtain
\begin{align}
\sum_{\alpha}c_{\alpha}Pc_{\alpha}^{\dagger}
&=
\kappa D\sqrt{\sigma}
\left(
\sum_{\alpha}p_{\alpha}
U_{\alpha}PU_{\alpha}^{\dagger}
\right)
\sqrt{\sigma}
\nonumber\\
&=
\kappa\sigma.
\end{align}
Taking the $i$th diagonal matrix element gives
\begin{equation}
\sum_{\alpha}
\langle i|c_{\alpha}Pc_{\alpha}^{\dagger}|i\rangle
=
\kappa\lambda_i.
\end{equation}

The unitary $1$-design condition also implies the adjoint twirling identity
\begin{equation}
\sum_{\alpha}p_{\alpha}
U_{\alpha}^{\dagger}XU_{\alpha}
=
\frac{\operatorname{Tr}X}{D}\mathbb{I}.
\end{equation}
Consequently,
\begin{align}
\sum_{\alpha}c_{\alpha}^{\dagger}c_{\alpha}
&=
\kappa D
\sum_{\alpha}p_{\alpha}
U_{\alpha}^{\dagger}\sigma U_{\alpha}
\nonumber\\
&=
\kappa \mathbb{I},
\end{align}
where we used $\operatorname{Tr}\sigma=1$.

Combining these two operator identities gives
\begin{align}
\mathcal{L}[P]
&=
\sum_{\alpha}
\left[
c_{\alpha}Pc_{\alpha}^{\dagger}
-
\frac{1}{2}
\left\{
c_{\alpha}^{\dagger}c_{\alpha},P
\right\}
\right]
\nonumber\\
&=
\kappa\sigma
-
\frac{\kappa}{2}\{\mathbb{I},P\}
\nonumber\\
&=
\kappa(\sigma-P).
\end{align}
Taking the $i$th diagonal matrix element, we obtain
\begin{equation}
\langle i|\mathcal{L}[P]|i\rangle
=
\kappa(\lambda_i-y_i).
\end{equation}

Averaging the stochastic equation for $P$ over the measurement outcomes,
and using $\mathbb{E}[dW_{\alpha}]=0$, gives
\begin{equation}
\frac{d\rho}{dt}
=
\mathcal{L}[\rho]
=
-\kappa(\rho-\sigma),
\qquad
\rho=\mathbb{E}[P].
\end{equation}

For the stochastic contribution, define
\begin{equation}
|\eta_{\alpha}\rangle
=
c_{\alpha}|\psi\rangle,
\qquad
\eta_{\alpha,i}
=
\langle i|\eta_{\alpha}\rangle.
\end{equation}
Using
\begin{equation}
M_{\alpha}(P)
=
c_{\alpha}P
+
Pc_{\alpha}^{\dagger}
-
O_{\alpha}(P)P,
\end{equation}
we obtain
\begin{align}
\langle i|M_{\alpha}(P)|i\rangle
&=
\langle i|c_{\alpha}|\psi\rangle
\langle\psi|i\rangle
+
\langle i|\psi\rangle
\langle\psi|c_{\alpha}^{\dagger}|i\rangle
-
O_{\alpha}(P)\langle i|P|i\rangle
\nonumber\\
&=
\psi_i^{*}\eta_{\alpha,i}
+
\eta_{\alpha,i}^{*}\psi_i
-
O_{\alpha}(P)y_i.
\end{align}

Hence the stochastic equation for the amplitude variables is
\begin{equation}
dy_i
=
\kappa(\lambda_i-y_i)\,dt
+
\sum_{\alpha}g_{i\alpha}(\psi)\,dW_{\alpha},
\label{SSE}
\end{equation}
with
\begin{equation}
g_{i\alpha}(\psi)
=
\psi_i^{*}\eta_{\alpha,i}
+
\eta_{\alpha,i}^{*}\psi_i
-
O_{\alpha}(P)y_i.
\end{equation}
For the independent coordinates, we take $i=1,\ldots,D-1$, while the
remaining variable is fixed by
\begin{equation}
y_D
=
1-\sum_{i=1}^{D-1}y_i.
\end{equation}

The first term, proportional to $dt$, represents a deterministic drift with velocity
\begin{equation}
v_i^{(a)}(y)=\kappa(\lambda_i-y_i).
\end{equation}
The diffusion of the amplitude variables caused by stochastic measurement outcomes is characterized by their
infinitesimal covariance matrix,
\begin{align}
D_{ij}^{(a)}(\bm y)
&\equiv
\frac{dy_i^{(d)}dy_j^{(d)}}{dt}
\nonumber\\
&=
\sum_{\alpha,\beta}
g_{i\alpha}(\psi)g_{j\beta}(\psi)
\frac{dW_{\alpha}dW_{\beta}}{dt}
\nonumber\\
&=
\sum_{\alpha}
g_{i\alpha}(\psi)g_{j\alpha}(\psi),
\end{align}
where $dy_i^{(d)}$ denotes the second term in Eq. (\ref{SSE}), and we have used the It\^o relation
\begin{equation}
dW_{\alpha}dW_{\beta}
=
\delta_{\alpha\beta}\,dt.
\end{equation}
Although each $g_{i\alpha}(\psi)$ generally depends on both the amplitudes
and phases of $|\psi\rangle$, we will show below that the sum over
measurement channels depends only on the amplitude variables $y$.

To evaluate the covariance matrix, we rewrite the noise coefficient as
\begin{equation}
g_{i\alpha}
=
h_{i\alpha}
+
h_{i\alpha}^*,
\end{equation}
where
\begin{equation}
h_{i\alpha}
=
\psi_i^*\eta_{\alpha,i}
-
y_i\langle\psi|c_\alpha|\psi\rangle 
= \langle \ell_i|c_\alpha|\psi\rangle ,\qquad \langle \ell_i|=\psi_i^*\langle i|-y_i \langle \psi|.
\end{equation}
Since \(\langle\ell_i|\psi\rangle=0\),
this vector lies in the tangent space of the projective Hilbert space and
represents the fluctuation direction of the coordinate
\(y_i\).

The covariance matrix can therefore be written as
\begin{align}
D_{ij}^{(a)}(\bm y)
&=
\sum_\alpha g_{i\alpha}g_{j\alpha}
\nonumber\\
&=
\sum_\alpha
\Big(
h_{i\alpha}h_{j\alpha}
+
h_{i\alpha}h_{j\alpha}^*
+
h_{i\alpha}^*h_{j\alpha}
+
h_{i\alpha}^*h_{j\alpha}^*
\Big).
\end{align}

The anomalous contractions vanish as a consequence of the phase-neutral
condition.  Indeed,
\begin{align}
\sum_\alpha c_\alpha\otimes c_\alpha
&=
\kappa D
\left(
\sqrt{\sigma}\otimes\sqrt{\sigma}
\right)
\sum_\alpha p_\alpha
U_\alpha\otimes U_\alpha
\nonumber\\
&=
0,
\end{align}
and the \(\ell_i\) representation gives
\[
\sum_\alpha h_{i\alpha}h_{j\alpha}
=
(\langle\ell_i|\otimes\langle\ell_j|)
\left(\sum_\alpha c_\alpha\otimes c_\alpha\right)
(|\psi\rangle\otimes|\psi\rangle).
\]
Consequently,
\begin{equation}
\sum_\alpha h_{i\alpha}h_{j\alpha}
=
\sum_\alpha h_{i\alpha}^*h_{j\alpha}^*
=
0.
\end{equation}

We next evaluate the normal contraction. The unitary $1$-design condition
gives
\begin{align}
\sum_\alpha
|\eta_\alpha\rangle\langle\eta_\alpha|
&=
\sum_\alpha
c_\alpha P c_\alpha^\dagger
\nonumber\\
&=
\kappa D\sqrt{\sigma}
\left(
\sum_\alpha p_\alpha
U_\alpha P U_\alpha^\dagger
\right)
\sqrt{\sigma}
\nonumber\\
&=
\kappa\sigma.
\label{eta_sum}
\end{align}
Using this identity, the normal contraction reduces to a single matrix
element:
\begin{align}
\sum_\alpha h_{i\alpha}h_{j\alpha}^*
&=
\langle\ell_i|
\left(
\sum_\alpha|\eta_\alpha\rangle\langle\eta_\alpha|
\right)
|\ell_j\rangle
\nonumber\\
&=
\kappa\langle\ell_i|\sigma|\ell_j\rangle
\nonumber\\
&=
\kappa
\Big[
\psi_i^*\psi_j\langle i|\sigma|j\rangle
-\psi_i^*y_j\langle i|\sigma|\psi\rangle
-y_i\psi_j\langle\psi|\sigma|j\rangle
+y_iy_j\langle\psi|\sigma|\psi\rangle
\Big]
\nonumber\\
&=
\kappa
\left[
\delta_{ij}\lambda_i y_i
+
\left(
s(\bm y)-\lambda_i-\lambda_j
\right)y_i y_j
\right],
\end{align}
where
\begin{equation}
s(\bm y)
=
\langle\psi|\sigma|\psi\rangle
=
\sum_{k=1}^D\lambda_k y_k.
\end{equation}
The result is real and symmetric under $i\leftrightarrow j$, and hence
\begin{equation}
\sum_\alpha h_{i\alpha}^*h_{j\alpha}
=
\sum_\alpha h_{i\alpha}h_{j\alpha}^*.
\end{equation}
Combining the normal and anomalous contractions, we finally obtain
\begin{equation}
D_{ij}^{(a)}(\bm y)
=
2\kappa
\left[
\delta_{ij}\lambda_i y_i
+
\left(
s(\bm y)-\lambda_i-\lambda_j
\right)y_i y_j
\right].
\end{equation}

We next derive the stochastic dynamics of the relative phase variables from the dynamics of the projector $P$. In the eigenbasis of $\sigma$, the
off-diagonal matrix element
\begin{equation}
P_{iD}
=
\psi_i\psi_D^*
=
\sqrt{y_i y_D}\,e^{i\varphi_i},
\qquad
\varphi_i=\theta_i-\theta_D,
\end{equation}
contains the relative phase between the $i$th and $D$th components.
Therefore, in the interior of the probability simplex, where
$y_i,y_D>0$, we may write
\begin{equation}
\varphi_i
=
\operatorname{Im}\ln P_{iD}.
\end{equation}

Taking the $(i,D)$ matrix element of
\begin{equation}
dP
=
\mathcal{L}(P)\,dt
+
\sum_\alpha M_\alpha(P)\,dW_\alpha,
\end{equation}
we obtain
\begin{equation}
dP_{iD}
=
\mathcal{L}(P)_{iD}\,dt
+
\sum_\alpha M_\alpha(P)_{iD}\,dW_\alpha.
\end{equation}
Applying the It\^o formula to $\ln P_{iD}$ gives
\begin{align}
d\ln P_{iD}
&=
\frac{dP_{iD}}{P_{iD}}
-
\frac{1}{2}
\frac{(dP_{iD})^2}{P_{iD}^2}
\nonumber\\
&=
\left[
\frac{\mathcal{L}(P)_{iD}}{P_{iD}}
-
\frac{1}{2}
\sum_\alpha
\left(
\frac{M_\alpha(P)_{iD}}{P_{iD}}
\right)^2
\right]dt
\nonumber\\
&\quad+
\sum_\alpha
\frac{M_\alpha(P)_{iD}}{P_{iD}}\,dW_\alpha.
\end{align}
Taking the imaginary part, the stochastic equation for $\varphi_i$ is
\begin{equation}
d\varphi_i
=
v_i^{(\varphi)}\,dt
+
\sum_\alpha q_{i\alpha}\,dW_\alpha,
\end{equation}
where
\begin{align}
v_i^{(\varphi)}
&=
\operatorname{Im}
\left[
\frac{\mathcal{L}(P)_{iD}}{P_{iD}}
-
\frac{1}{2}
\sum_\alpha
\left(
\frac{M_\alpha(P)_{iD}}{P_{iD}}
\right)^2
\right],\\
q_{i\alpha}
&=
\operatorname{Im}
\left[
\frac{M_\alpha(P)_{iD}}{P_{iD}}
\right].
\end{align}

We first evaluate the deterministic contribution. Since $\sigma$ is diagonal in the chosen basis, for $i\neq D$ we have
\begin{align}
\mathcal{L}(P)_{iD}
&=
\left(
\kappa\sigma-\kappa P
\right)_{iD}
\nonumber\\
&=
-\kappa P_{iD}.
\end{align}
Consequently,
\begin{equation}
\frac{\mathcal{L}(P)_{iD}}{P_{iD}}
=
-\kappa,
\end{equation}
which is purely real and does not contribute to the phase drift.

For the stochastic contribution, using
\begin{equation}
M_\alpha(P)
=
c_\alpha P
+
Pc_\alpha^\dagger
-
O_\alpha P
\end{equation}
and $|\eta_\alpha\rangle=c_\alpha|\psi\rangle$, we find
\begin{align}
\frac{M_\alpha(P)_{iD}}{P_{iD}}
&=
\frac{\eta_{\alpha,i}}{\psi_i}
+
\frac{\eta_{\alpha,D}^*}{\psi_D^*}
-
O_\alpha\nonumber\\
&=
\left(
\frac{\eta_{\alpha,i}}{\psi_i}
-
\langle\psi|c_\alpha|\psi\rangle
\right)
+
\left(
\frac{\eta_{\alpha,D}^*}{\psi_D^*}
-
\langle\psi|c_\alpha^\dagger|\psi\rangle
\right).
\end{align}

Squaring this expression and summing over $\alpha$, we obtain

\begin{align}
\sum_\alpha
\left(
\frac{M_\alpha(P)_{iD}}{P_{iD}}
\right)^2
&=
\sum_\alpha
\left(
\frac{\eta_{\alpha,i}}{\psi_i}
-
\langle\psi|c_\alpha|\psi\rangle
\right)^2
\nonumber\\
&\quad+
\sum_\alpha
\left(
\frac{\eta_{\alpha,D}^*}{\psi_D^*}
-
\langle\psi|c_\alpha^\dagger|\psi\rangle
\right)^2
\nonumber\\
&\quad+
2\sum_\alpha
\left(
\frac{\eta_{\alpha,i}}{\psi_i}
-
\langle\psi|c_\alpha|\psi\rangle
\right)
\left(
\frac{\eta_{\alpha,D}^*}{\psi_D^*}
-
\langle\psi|c_\alpha^\dagger|\psi\rangle
\right).
\end{align}

The first two terms are anomalous contractions. They vanish as a consequence
of the phase-neutral condition
\begin{equation}
\sum_\alpha c_\alpha\otimes c_\alpha=0.
\end{equation}

The remaining normal contraction is
\begin{align}
&\sum_\alpha
\left(
\frac{\eta_{\alpha,i}}{\psi_i}
-
\langle\psi|c_\alpha|\psi\rangle
\right)
\left(
\frac{\eta_{\alpha,D}^*}{\psi_D^*}
-
\langle\psi|c_\alpha^\dagger|\psi\rangle
\right)
\nonumber\\
&=
\sum_\alpha
\frac{\eta_{\alpha,i}\eta_{\alpha,D}^*}
{\psi_i\psi_D^*}
-
\sum_\alpha
\frac{\eta_{\alpha,i}}{\psi_i}
\langle\psi|c_\alpha^\dagger|\psi\rangle
\nonumber\\
&\quad
-
\sum_\alpha
\langle\psi|c_\alpha|\psi\rangle
\frac{\eta_{\alpha,D}^*}{\psi_D^*}
+
\sum_\alpha
\left|
\langle\psi|c_\alpha|\psi\rangle
\right|^2.
\end{align}
Using Eq. (\ref{eta_sum}) together with $i\neq D$, the four terms are
\begin{align}
\sum_\alpha
\frac{\eta_{\alpha,i}\eta_{\alpha,D}^*}
{\psi_i\psi_D^*}
&=0,\\
\sum_\alpha
\frac{\eta_{\alpha,i}}{\psi_i}
\langle\psi|c_\alpha^\dagger|\psi\rangle
&=\kappa\lambda_i,\\
\sum_\alpha
\langle\psi|c_\alpha|\psi\rangle
\frac{\eta_{\alpha,D}^*}{\psi_D^*}
&=\kappa\lambda_D,\\
\sum_\alpha
\left|
\langle\psi|c_\alpha|\psi\rangle
\right|^2
&=\kappa s(\bm y).
\end{align}
Therefore,
\begin{align}
&\sum_\alpha
\left(
\frac{\eta_{\alpha,i}}{\psi_i}
-
\langle\psi|c_\alpha|\psi\rangle
\right)
\left(
\frac{\eta_{\alpha,D}^*}{\psi_D^*}
-
\langle\psi|c_\alpha^\dagger|\psi\rangle
\right)
\nonumber\\
&\qquad=
\kappa
\left[
s(\bm y)-\lambda_i-\lambda_D
\right].
\end{align}
Combining the anomalous and normal contractions, we obtain
\begin{equation}
\sum_\alpha
\left(
\frac{M_\alpha(P)_{iD}}{P_{iD}}
\right)^2
=
2\kappa
\left[
s(\bm y)-\lambda_i-\lambda_D
\right].
\end{equation}
This quantity is real and therefore does not contribute to the phase drift.
Therefore,
\begin{equation}
v_i^{(\varphi)}=0.
\end{equation}

Since $O_\alpha$ is real, the noise coefficient becomes
\begin{equation}
q_{i\alpha}
=
\operatorname{Im}
\left(
\frac{\eta_{\alpha,i}}{\psi_i}
-
\frac{\eta_{\alpha,D}}{\psi_D}
\right).
\end{equation}

Their infinitesimal covariance matrix is
\begin{equation}
D_{ij}^{(\varphi)}(y)
=
\sum_\alpha q_{i\alpha}q_{j\alpha}.
\end{equation}
Using
\begin{equation}
\operatorname{Im}u\,\operatorname{Im}v
=
\frac{1}{2}
\operatorname{Re}
\left(
uv^*-uv
\right),
\end{equation}
we obtain
\begin{align}
D_{ij}^{(\varphi)}(y)
&=
\frac{1}{2}
\operatorname{Re}
\sum_\alpha
\Bigg[
\left(
\frac{\eta_{\alpha,i}}{\psi_i}
-
\frac{\eta_{\alpha,D}}{\psi_D}
\right)
\left(
\frac{\eta_{\alpha,j}^*}{\psi_j^*}
-
\frac{\eta_{\alpha,D}^*}{\psi_D^*}
\right)
\nonumber\\
&\hspace{3.2cm}
-
\left(
\frac{\eta_{\alpha,i}}{\psi_i}
-
\frac{\eta_{\alpha,D}}{\psi_D}
\right)
\left(
\frac{\eta_{\alpha,j}}{\psi_j}
-
\frac{\eta_{\alpha,D}}{\psi_D}
\right)
\Bigg].
\end{align}
The second term vanishes by the phase-neutral condition. For the first
term, the unitary $1$-design condition gives
\begin{equation}
\sum_\alpha
\eta_{\alpha,i}\eta_{\alpha,j}^*
=
\kappa\lambda_i\delta_{ij}.
\end{equation}
Therefore,
\begin{equation}
D_{ij}^{(\varphi)}(y)
=
\frac{\kappa}{2}
\left(
\delta_{ij}\frac{\lambda_i}{y_i}
+
\frac{\lambda_D}{y_D}
\right),
\qquad
i,j=1,\ldots,D-1.
\end{equation}

Collecting the results derived above, the stochastic dynamics of the
coordinates $\bm z=(\bm y,\bm\varphi)$ can be written as
\begin{equation}
dz_\mu
=
v_\mu(\bm z)\,dt
+
\sum_\alpha
G_{\mu\alpha}(\bm z)\,dW_\alpha,
\end{equation}
where $\mu=1,\ldots,2(D-1)$.
The drift vector is
\begin{equation}
v_\mu(\bm z)
=
\left(
v_i^{(a)}(\bm y),0
\right),
\qquad
v_i^{(a)}(\bm y)
=
\kappa(\lambda_i-y_i),
\end{equation}
where $i=1,\ldots,D-1$. The corresponding noise coefficients are defined by
\begin{equation}
G_{\mu\alpha}(\bm z)
=
\begin{cases}
g_{i\alpha}(\psi),
& \mu=i,\\[2mm]
q_{i\alpha}(\psi),
& \mu=D-1+i,
\end{cases}
\qquad
i=1,\ldots,D-1.
\end{equation}
The covariance matrix of the full stochastic process is then
\begin{equation}
D_{\mu\nu}(\bm z)
=
\sum_\alpha
G_{\mu\alpha}(\bm z)
G_{\nu\alpha}(\bm z).
\end{equation}

It remains to evaluate the mixed covariance between the amplitude and
relative phase variables,
\begin{equation}
D_{ij}^{(a\varphi)}(\bm z)
=
\sum_\alpha
g_{i\alpha}q_{j\alpha}.
\end{equation}
Using
\begin{equation}
g_{i\alpha}
=
h_{i\alpha}+h_{i\alpha}^*,
\qquad
q_{j\alpha}
=
\operatorname{Im}
\left(
\frac{\eta_{\alpha,j}}{\psi_j}
-
\frac{\eta_{\alpha,D}}{\psi_D}
\right),
\end{equation}
we obtain
\begin{align}
D_{ij}^{(a\varphi)}
&=
\operatorname{Im}
\sum_\alpha
\left(
h_{i\alpha}+h_{i\alpha}^*
\right)
\left(
\frac{\eta_{\alpha,j}}{\psi_j}
-
\frac{\eta_{\alpha,D}}{\psi_D}
\right)
\nonumber\\
&=
-\operatorname{Im}
\sum_\alpha
h_{i\alpha}
\left(
\frac{\eta_{\alpha,j}^*}{\psi_j^*}
-
\frac{\eta_{\alpha,D}^*}{\psi_D^*}
\right).
\end{align}
Here, the anomalous contractions vanish by the phase-neutral condition.

Using
\begin{equation}
h_{i\alpha}
=
\psi_i^*\eta_{\alpha,i}
-
y_i\langle\psi|c_\alpha|\psi\rangle
\end{equation}
and
\begin{equation}
\sum_\alpha
|\eta_\alpha\rangle\langle\eta_\alpha|
=
\kappa\sigma,
\end{equation}
the remaining normal contraction is
\begin{align}
&\sum_\alpha
h_{i\alpha}
\left(
\frac{\eta_{\alpha,j}^*}{\psi_j^*}
-
\frac{\eta_{\alpha,D}^*}{\psi_D^*}
\right)
\nonumber\\
&=
\kappa
\left[
\delta_{ij}\lambda_i
-
y_i\lambda_j
+
y_i\lambda_D
\right].
\end{align}
This expression is real. Therefore,
\begin{equation}
D_{ij}^{(a\varphi)}(\bm z)=0.
\end{equation}
The covariance matrix consequently takes the block-diagonal form
\begin{equation}
D_{\mu\nu}(\bm z)
=
\begin{pmatrix}
D_{ij}^{(a)}(\bm y) & 0\\
0 & D_{ij}^{(\varphi)}(\bm y)
\end{pmatrix}.
\end{equation}

The two blocks are
\begin{equation}
D_{ij}^{(a)}(\bm y)
=
2\kappa
\left[
\delta_{ij}\lambda_i y_i
+
\left(
s(\bm y)-\lambda_i-\lambda_j
\right)y_i y_j
\right],
\end{equation}
and
\begin{equation}
D_{ij}^{(\varphi)}(\bm y)
=
\frac{\kappa}{2}
\left(
\delta_{ij}\frac{\lambda_i}{y_i}
+
\frac{\lambda_D}{y_D}
\right),
\end{equation}
where
\begin{equation}
s(\bm y)
=
\sum_{k=1}^D\lambda_k y_k.
\end{equation}

Accordingly, the probability density $p(\bm z,t)$ associated with this stochastic process obeys the standard Fokker--Planck equation

\begin{equation}
\frac{\partial p}{\partial t}
=
-\sum_\mu
\partial_\mu
\left(
v_\mu p
\right)
+
\frac{1}{2}
\sum_{\mu,\nu}
\partial_\mu\partial_\nu
\left(
D_{\mu\nu}p
\right).
\end{equation}

\section{Proof of the main theorem}

We now prove the main theorem. The proof consists of two steps. First, we
show that the Scrooge distribution derived above is a stationary solution of
the Fokker--Planck equation. We then establish the uniqueness of this
stationary distribution.

The Scrooge probability density is
\begin{equation}
p_{\mathrm{sc}}(\bm y,\bm\varphi)
=
C H(\bm y)^{-(D+1)},
\qquad
H(\bm y)
=
\sum_{k=1}^D\frac{y_k}{\lambda_k}.
\end{equation}
It is independent of the relative phase variables $\bm\varphi$. Since
$D_{ij}^{(\varphi)}(\bm y)$ is also independent of $\bm\varphi$, the phase
diffusion term in the Fokker--Planck equation vanishes:
\begin{equation}
\sum_{i,j=1}^{D-1}
\frac{\partial^2}{\partial\varphi_i\partial\varphi_j}
\left[
D_{ij}^{(\varphi)}(\bm y)
p_{\mathrm{sc}}(\bm y,\bm\varphi)
\right]
=
0.
\end{equation}

It therefore
remains to consider the amplitude probability current
\begin{equation}
J_i^{(a)}
=
v_i^{(a)}p_{\mathrm{sc}}
-
\frac{1}{2}
\sum_{j=1}^{D-1}
\partial_j
\left(
D_{ij}^{(a)}p_{\mathrm{sc}}
\right).
\end{equation}
We will show that this current vanishes pointwise.

Since
\begin{equation}
p_{\mathrm{sc}}
=
C H(\bm y)^{-(D+1)},
\end{equation}
and
\begin{equation}
\partial_jH
=
\frac{1}{\lambda_j}
-
\frac{1}{\lambda_D},
\end{equation}
we have
\begin{equation}
\partial_jp_{\mathrm{sc}}
=
-\frac{D+1}{H(\bm y)}
\left(
\frac{1}{\lambda_j}
-
\frac{1}{\lambda_D}
\right)
p_{\mathrm{sc}}.
\end{equation}

Recall that
\begin{equation}
D_{ij}^{(a)}
=
2\kappa
\left[
\delta_{ij}\lambda_i y_i
+
\left(
s(\bm y)-\lambda_i-\lambda_j
\right)y_i y_j
\right].
\end{equation}
Using
\begin{equation}
\partial_jy_i=\delta_{ij},
\qquad
\partial_js(\bm y)=\lambda_j-\lambda_D,
\end{equation}
and summing over the independent variables $y_1,\ldots,y_{D-1}$, we obtain
\begin{equation}
\sum_{j=1}^{D-1}
\partial_jD_{ij}^{(a)}
=
2\kappa
\left\{
\lambda_i
+
y_i
\left[
(D+1)
\left(
s(\bm y)-\lambda_i
\right)
-
1
\right]
\right\}.
\end{equation}

Substituting the derivative of $p_{\mathrm{sc}}$, we obtain
\begin{align}
\sum_{j=1}^{D-1}
D_{ij}^{(a)}
\partial_jp_{\mathrm{sc}}
=
-2\kappa(D+1)y_i
\left(
s(\bm y)-\lambda_i
\right)
p_{\mathrm{sc}}.
\end{align}

Expanding the derivative in the probability current gives
\begin{align}
J_i^{(a)}
&=
v_i^{(a)}p_{\mathrm{sc}}
-
\frac{1}{2}
p_{\mathrm{sc}}
\sum_{j=1}^{D-1}
\partial_jD_{ij}^{(a)}
-
\frac{1}{2}
\sum_{j=1}^{D-1}
D_{ij}^{(a)}
\partial_jp_{\mathrm{sc}}.
\end{align}
Substituting the expressions above and
$v_i^{(a)}=\kappa(\lambda_i-y_i)$, we find
\begin{align}
\frac{J_i^{(a)}}{\kappa p_{\mathrm{sc}}}
&=
\lambda_i-y_i
-
\lambda_i
-
y_i
\left[
(D+1)
\left(
s(\bm y)-\lambda_i
\right)
-
1
\right]
\nonumber\\
&\quad+
(D+1)y_i
\left(
s(\bm y)-\lambda_i
\right)
\nonumber\\
&=
0.
\end{align}
Therefore,
\begin{equation}
J_i^{(a)}=0,
\qquad
i=1,\ldots,D-1.
\end{equation}
All terms on the right-hand side of the Fokker--Planck equation thus vanish
when $p=p_{\mathrm{sc}}$, and hence
\begin{equation}
\left.
\frac{\partial p}{\partial t}
\right|_{p=p_{\mathrm{sc}}}
=
0.
\end{equation}
This proves that the Scrooge distribution is a stationary distribution of
the stochastic dynamics.

We now establish the uniqueness of the stationary distribution and
convergence toward it. Our strategy is to verify the \(t_0\)-regularity
required by Doob's theorem through the uniform ellipticity of the
diffusion on the projective Hilbert space.

Let \(X_t\) denote the pure-state Markov process on
\begin{equation}
\mathcal{M}=\mathbb{CP}^{D-1},
\end{equation}
and let
\begin{equation}
\mathcal{P}_t(x,A)
=
\mathbb{P}\!\left(X_t\in A\,\middle|\,X_0=x\right)
\end{equation}
denote its transition probability. For an initial probability measure
\(\nu\), we define
\begin{equation}
(\nu\mathcal{P}_t)(A)
=
\int_{\mathcal{M}} \mathcal{P}_t(x,A)\,\nu(dx).
\end{equation}

\begin{lemma}[Doob's theorem~\cite{KulikScheutzow2015,DaPratoZabczyk1996}]
Let \((\mathcal{P}_t)_{t\geq0}\) be a stochastically continuous Markov
semigroup on a Polish space. Suppose that it admits an invariant
probability measure \(\mu\) and that, for some \(t_0>0\),
\begin{equation}
\mathcal{P}_{t_0}(x,\cdot)
\sim
\mathcal{P}_{t_0}(x',\cdot)
\end{equation}
for every pair of states \(x,x'\), where \(\sim\) denotes mutual absolute
continuity, namely, the two probability measures have the same null sets.
Then \(\mu\) is the unique invariant probability measure and
\begin{equation}
\left\|
\mathcal{P}_t(x,\cdot)-\mu
\right\|_{\mathrm{TV}}
\longrightarrow 0,
\qquad
t\to\infty,
\end{equation}
for every \(x\). Here, \(\|\cdot\|_{\mathrm{TV}}\) denotes the total
variation distance,
\begin{equation}
\|\chi-\nu\|_{\mathrm{TV}}
=
\sup_A
|\chi(A)-\nu(A)|,
\end{equation}
which measures the largest difference between the probabilities assigned
by \(\chi\) and \(\nu\) to the same measurable event.
\end{lemma}

\begin{lemma}[Uniform ellipticity]
Suppose that \(\sigma\) is full rank and define
\begin{equation}
\lambda_{\min}
=
\min_{1\leq k\leq D}\lambda_k
>0.
\end{equation}
Let \(g_{\mathrm{FS}}\) denote the Fubini--Study metric on
\(\mathbb{CP}^{D-1}\). Then, for every
\(x\in\mathbb{CP}^{D-1}\) and every cotangent vector
\(\xi\in T_x^*\mathbb{CP}^{D-1}\),
\begin{equation}
D_x(\xi,\xi)
\geq
\frac{\kappa\lambda_{\min}}{2}
g_{\mathrm{FS},x}^{-1}(\xi,\xi).
\end{equation}
Hence the diffusion is uniformly elliptic.
\end{lemma}

\begin{proof}
The covariance matrix derived above defines a contravariant diffusion
tensor acting on cotangent vectors through
\begin{equation}
D_x(\xi,\xi)
=
\sum_{\mu,\nu}
D_{\mu\nu}(x)\xi_\mu\xi_\nu.
\end{equation}

Write
\begin{equation}
\psi_k
=
\sqrt{y_k}\,e^{i\theta_k},
\qquad
\sum_{k=1}^{D}y_k=1.
\end{equation}
The Fubini--Study line element is
\begin{equation}
ds_{\mathrm{FS}}^2
=
\langle d\psi|d\psi\rangle
-
\left|\langle\psi|d\psi\rangle\right|^2.
\end{equation}
Since
\begin{equation}
d\psi_k
=
e^{i\theta_k}
\left(
\frac{dy_k}{2\sqrt{y_k}}
+
i\sqrt{y_k}\,d\theta_k
\right),
\end{equation}
we obtain
\begin{equation}
ds_{\mathrm{FS}}^2
=
\frac{1}{4}
\sum_{k=1}^{D}\frac{dy_k^2}{y_k}
+
\sum_{k=1}^{D}y_kd\theta_k^2
-
\left(
\sum_{k=1}^{D}y_kd\theta_k
\right)^2.
\end{equation}

Using the independent coordinates
\begin{equation}
y_D
=
1-\sum_{i=1}^{D-1}y_i,
\qquad
\varphi_i
=
\theta_i-\theta_D,
\end{equation}
the metric is block diagonal:
\begin{equation}
ds_{\mathrm{FS}}^2
=
ds_{(a)}^2
+
ds_{(\varphi)}^2,
\end{equation}
where
\begin{equation}
ds_{(a)}^2
=
\frac{1}{4}
\left[
\sum_{i=1}^{D-1}\frac{dy_i^2}{y_i}
+
\frac{
\left(\sum_{i=1}^{D-1}dy_i\right)^2
}{y_D}
\right],
\end{equation}
and
\begin{equation}
ds_{(\varphi)}^2
=
\sum_{i=1}^{D-1}y_i\,d\varphi_i^2
-
\left(
\sum_{i=1}^{D-1}y_i\,d\varphi_i
\right)^2.
\end{equation}

Introduce
\begin{equation}
Y
=
\operatorname{diag}(y_1,\ldots,y_{D-1}),
\qquad
\bm y
=
(y_1,\ldots,y_{D-1})^T,
\qquad
\bm 1
=
(1,\ldots,1)^T\in\mathbb{R}^{D-1}.
\end{equation}
The two metric blocks are
\begin{equation}
g_{\mathrm{FS}}^{(a)}
=
\frac{1}{4}
\left(
Y^{-1}
+
\frac{1}{y_D}\bm 1\bm 1^T
\right),
\qquad
g_{\mathrm{FS}}^{(\varphi)}
=
Y-\bm y\bm y^T.
\end{equation}
The Sherman--Morrison formula gives
\begin{equation}
\left(g_{\mathrm{FS}}^{(a)}\right)^{-1}
=
4\left(
Y-\bm y\bm y^T
\right),
\end{equation}
and
\begin{equation}
\left(g_{\mathrm{FS}}^{(\varphi)}\right)^{-1}
=
Y^{-1}
+
\frac{1}{y_D}\bm 1\bm 1^T.
\end{equation}

Let
\begin{equation}
\xi
=
\sum_{i=1}^{D-1}a_i\,dy_i
+
\sum_{i=1}^{D-1}b_i\,d\varphi_i
\end{equation}
be an arbitrary cotangent vector. Set
\begin{equation}
a_D=0,
\qquad
\bar a
=
\sum_{k=1}^{D}y_ka_k.
\end{equation}
The amplitude part of the inverse metric is
\begin{align}
\left(g_{\mathrm{FS}}^{(a)}\right)^{-1}(a,a)
&=
4
\left[
\sum_{i=1}^{D-1}y_i a_i^2
-
\left(
\sum_{i=1}^{D-1}y_i a_i
\right)^2
\right]
\nonumber\\
&=
4
\sum_{k=1}^{D}
y_k(a_k-\bar a)^2.
\end{align}
The phase part is
\begin{equation}
\left(g_{\mathrm{FS}}^{(\varphi)}\right)^{-1}(b,b)
=
\sum_{i=1}^{D-1}\frac{b_i^2}{y_i}
+
\frac{1}{y_D}
\left(
\sum_{i=1}^{D-1}b_i
\right)^2.
\end{equation}

Using the amplitude diffusion block derived above, we have
\begin{equation}
D^{(a)}(a,a)
=
2\kappa
\sum_{k=1}^{D}
\lambda_k y_k(a_k-\bar a)^2.
\end{equation}
Therefore,
\begin{align}
D^{(a)}(a,a)
&\geq
2\kappa\lambda_{\min}
\sum_{k=1}^{D}
y_k(a_k-\bar a)^2
\nonumber\\
&=
\frac{\kappa\lambda_{\min}}{2}
\left(g_{\mathrm{FS}}^{(a)}\right)^{-1}(a,a).
\end{align}

Similarly, the phase diffusion block gives
\begin{equation}
D^{(\varphi)}(b,b)
=
\frac{\kappa}{2}
\left[
\sum_{i=1}^{D-1}
\frac{\lambda_i}{y_i}b_i^2
+
\frac{\lambda_D}{y_D}
\left(
\sum_{i=1}^{D-1}b_i
\right)^2
\right].
\end{equation}
Hence
\begin{equation}
D^{(\varphi)}(b,b)
\geq
\frac{\kappa\lambda_{\min}}{2}
\left(g_{\mathrm{FS}}^{(\varphi)}\right)^{-1}(b,b).
\end{equation}

Since both the Fubini--Study metric and the diffusion tensor are block
diagonal in these coordinates,
\begin{align}
D_x(\xi,\xi)
&=
D^{(a)}(a,a)
+
D^{(\varphi)}(b,b)
\nonumber\\
&\geq
\frac{\kappa\lambda_{\min}}{2}
\left[
\left(g_{\mathrm{FS}}^{(a)}\right)^{-1}(a,a)
+
\left(g_{\mathrm{FS}}^{(\varphi)}\right)^{-1}(b,b)
\right]
\nonumber\\
&=
\frac{\kappa\lambda_{\min}}{2}
g_{\mathrm{FS},x}^{-1}(\xi,\xi).
\end{align}

The set on which all \(y_k>0\) is dense in
\(\mathbb{CP}^{D-1}\). Since both sides are smooth tensorial quadratic
forms, the inequality extends by continuity to the whole manifold.
This proves uniform ellipticity.
\end{proof}

\begin{lemma}[\(t_0\)-regularity~\cite{SaloffCoste1992}]
Let \(\mathcal{M}\) be a connected compact smooth manifold, and let
\((\mathcal{P}_t)_{t\geq0}\) be a diffusion process on \(\mathcal{M}\)
whose generator has smooth coefficients and is uniformly elliptic with
respect to a Riemannian metric \(g\). Then, for every \(t>0\),
\(\mathcal{P}_t(x,\cdot)\) has a smooth and strictly positive density
\(p_t(x,y)\) with respect to the Riemannian volume measure
\(d\operatorname{vol}_g(y)\). Consequently,
\begin{equation}
\mathcal{P}_t(x,\cdot)
\sim
\mathcal{P}_t(x',\cdot)
\end{equation}
for all \(x,x'\in\mathcal{M}\), and the process is \(t_0\)-regular for
every \(t_0>0\).
\end{lemma}

\begin{proof}
In local coordinates, the smooth diffusion generator can be written in
divergence form with smooth lower-order coefficients. Since
\(\mathcal{M}\) is compact, these coefficients are bounded.

Standard parabolic regularity implies that, for every \(t>0\), the
transition probability admits a smooth density:
\begin{equation}
\mathcal{P}_t(x,A)
=
\int_A
p_t(x,y)\,
d\operatorname{vol}_g(y).
\end{equation}

Uniform ellipticity gives a strictly positive local lower bound for the
fundamental solution. Since \(\mathcal{M}\) is connected, any two points
can be joined by a path covered by finitely many overlapping coordinate
neighborhoods. Applying the local lower bound successively and using
the Chapman--Kolmogorov relation gives
\begin{equation}
p_t(x,y)>0
\end{equation}
for every \(t>0\) and all \(x,y\in\mathcal{M}\).

Therefore,
\begin{equation}
\mathcal{P}_t(x,A)=0
\quad\Longleftrightarrow\quad
\operatorname{vol}_g(A)=0,
\end{equation}
independently of \(x\). Hence the measures
\(\mathcal{P}_t(x,\cdot)\) are mutually equivalent for all \(x\), and
the process is \(t_0\)-regular for every \(t_0>0\).
\end{proof}

\begin{theorem}[Uniqueness and convergence to the Scrooge ensemble]
Suppose that \(\sigma\) is full rank. With a slight abuse of notation, \(p_{\mathrm{sc}}\) is the
unique stationary probability measure of the stochastic dynamics.
Moreover, for every initial probability measure \(\nu\),
\begin{equation}
\left\|
\nu\mathcal{P}_t-p_{\mathrm{sc}}
\right\|_{\mathrm{TV}}
\longrightarrow 0,
\qquad
t\to\infty.
\end{equation}
\end{theorem}

\begin{proof}
The Scrooge distribution has already been shown to be stationary.
The stochastic vector fields appearing in the stochastic Schr\"odinger
equation are smooth functions of the pure state, so the generator has
smooth coefficients.

The space \(\mathbb{CP}^{D-1}\) is compact and metrizable, and hence is
a Polish space. Moreover, the process has continuous sample paths, so
its Markov semigroup is stochastically continuous.

Uniform ellipticity follows from the second lemma, and the third lemma
then establishes \(t_0\)-regularity. Doob's theorem therefore implies
that \(p_{\mathrm{sc}}\) is the unique stationary probability measure and
that
\begin{equation}
\left\|
\mathcal{P}_t(x,\cdot)-p_{\mathrm{sc}}
\right\|_{\mathrm{TV}}
\longrightarrow 0
\end{equation}
for every initial state \(x\).

For an arbitrary initial probability measure \(\nu\),
\begin{equation}
\left\|
\nu\mathcal{P}_t-p_{\mathrm{sc}}
\right\|_{\mathrm{TV}}
\leq
\int_{\mathcal{M}}
\left\|
\mathcal{P}_t(x,\cdot)-p_{\mathrm{sc}}
\right\|_{\mathrm{TV}}
\,\nu(dx).
\end{equation}
The result then follows from dominated convergence.
\end{proof}

If \(\sigma\) is not full rank, let \(\Pi\) denote the projector onto
\(\operatorname{supp}\sigma\) and let \(r=\operatorname{rank}\sigma\).
Since
\begin{equation}
\mathbb{E}\!\left[
\operatorname{Tr}\!\left((\mathbb{I}-\Pi)P(t)\right)
\right]
=
e^{-\kappa t}
\mathbb{E}\!\left[
\operatorname{Tr}\!\left((\mathbb{I}-\Pi)P(0)\right)
\right],
\qquad
P(t)=|\psi_t\rangle\langle\psi_t|,
\end{equation}
any stationary probability measure must be supported on
\(\mathbb{CP}(\operatorname{supp}\sigma)\simeq\mathbb{CP}^{r-1}\).
This subspace is invariant because, for \(P=\Pi P\Pi\),
\(\mathcal{L}(P)=\Pi\mathcal{L}(P)\Pi\) and
\(M_\alpha(P)=\Pi M_\alpha(P)\Pi\).
Restricted to this subspace, \(\sigma\) is full rank, so the preceding
theorem implies that the corresponding Scrooge distribution is the
unique stationary measure and attracts every initial distribution
supported there in total variation. For general initial conditions, the
population outside \(\operatorname{supp}\sigma\) decays exponentially in
expectation, so the long-time dynamics asymptotically reduces to this
lower-dimensional projective space.

\bibliography{ref.bib}